\documentclass[12pt,draftcls,onecolumn]{IEEEtran}

\usepackage{amsfonts}
\usepackage{amssymb}
\usepackage{cite}
\usepackage[cmex10]{amsmath}
\usepackage[]{algorithm}
\usepackage{algorithmic}
\usepackage{subeqnarray}
\usepackage{cases}
\usepackage{graphicx}
\usepackage{subfigure}
\usepackage{booktabs}
\usepackage{xcolor}
\usepackage{soul}
\usepackage{multirow}
\usepackage{diagbox}

\ifCLASSINFOpdf
\else
\fi

\begin{document}
\title{CSI-PPPNet: A One-Sided One-for-All Deep Learning Framework for Massive MIMO CSI Feedback}
%
%
%

\author{
Wei~Chen,~\IEEEmembership{Senior~Member,~IEEE},~Weixiao~Wan,~Shiyue~Wang,~Peng~Sun,\\~Geoffrey~Ye~Li,~\IEEEmembership{Fellow,~IEEE},~Bo~Ai,~\IEEEmembership{Fellow,~IEEE}

\thanks{Wei Chen, Weixiao Wan, Shiyue Wang and Bo Ai are with the State Key Laboratory of Rail Traffic Control and Safety, Beijing Jiaotong University, Beijing, China (e-mail: weich, weixiaowan, shiyuewang, boai@bjtu.edu.cn).

Peng Sun is with vivo Communication Research Institute, Beijing 100015, P.R.China (Email: sunpeng@vivo.com).

Geoffrey Ye Li is with the Department of Electrical and Electronic Engineering, Imperial College London, London, U.K (Email: geoffrey.li@imperial.ac.uk).
}
}

\maketitle
\begin{abstract}
To reduce multiuser interference and maximize the spectrum efficiency in orthogonal frequency division duplexing massive multiple-input multiple-output (MIMO) systems, the downlink channel state information (CSI) estimated at the user equipment (UE) is required at the base station (BS). This paper presents a novel method for massive MIMO CSI feedback via a one-sided one-for-all deep learning framework. The CSI is compressed via linear projections at the UE, and is recovered at the BS using deep learning (DL) with plug-and-play priors (PPP). Instead of using handcrafted regularizers for the wireless channel responses, the proposed approach, namely CSI-PPPNet, exploits a DL based denoisor in place of the proximal operator of the prior in an alternating optimization scheme. In this way, a DL model trained once for denoising can be repurposed for CSI recovery tasks with arbitrary compression ratio. The one-sided one-for-all framework reduces model storage space, relieves the burden of joint model training and model delivery, and could be applied at UEs with limited device memories and computation power. Extensive experiments over the open indoor and urban macro scenarios show the effectiveness and advantages of the proposed method.
\end{abstract}


\IEEEpeerreviewmaketitle

\section{Introduction}
\IEEEPARstart{M}{assive} multiple-input multiple-output (MIMO) technology \cite{6736761} is an essential building block in 5G wireless communication systems and 5G-Advanced systems. Compared to MIMO technology used in 4G LTE networks, increasing the number of antennas improves the network capacity, leading to higher throughput and multiuser support, while it brings new challenges to the design of the communication system, e.g., channel information feedback, pilot signal design, codebook design and beam management \cite{8861014}.

\par
Through continuous evolution releases, the current 5G protocol supports various scenarios for operations with massive number of antennas \cite{3GPP5}. The major benefit of massive MIMO operation lies in multiuser operation to improve spectral efficiency in addition to the improved signal-to-interference-and-noise ratio (SINR) for single user operation. To realize these promised benefits, channel state information (CSI) is required \cite{8861014}. In the time-division duplexing (TDD) system, the base station (BS) estimates the uplink CSI through the pilot signals transmitted by the user equipment (UE), and then the downlink CSI can be inferred from the uplink CSI by using channel reciprocity. In the frequency-division duplex (FDD) system, channel reciprocity is no longer satisfied, as different frequencies are employed in the uplink and downlink. Thus, UEs are required to estimate their own downlink CSI through the local channel estimator, and then feed back the estimated CSI to the BS for subsequent usages. CSI feedback consumes spectrum and energy resources. The amount of CSI tends to grow with the number of antennas equipped at the BS. Therefore, it is usually desired to feed back a compressed version of the full CSI or features that are sufficient for subsequent usages, e.g., precoding designs. In the 5G New Radio (NR) networks, the Type I codebook and the Type II codebook are exploited for CSI feedback. The Type I codebook designed for single-user transmission is relatively simple, but has limited accuracy. The Type II codebook designed for multiuser scheduling has high precision, while the overhead of the Type II codebook is too high to capture fine-grain spatial channel structures in the massive MIMO case \cite{3GPP4}.


\par
As the massive MIMO channel tends to be sparse owing to the limited local scatters at the BS, various compressive sensing (CS) based CSI compression methods have been studied to address the challenge of overwhelming feedback in massive MIMO systems. The UE takes a reduced number of random measurements of the CSI, and reports the limited measurements to the BS. The original CSI could be recovered by exploiting the CS algorithms in the BS. CS is used in \cite{6214417} to reduce the CSI feedback load in a massive MIMO system by exploiting the expected sparsity in the spatial-frequency domain resulting from spatially correlated antenna arrays. The massive MIMO CSI feedback in \cite{6966062} uses multidimensional CS with Tucker tensor decomposition, which exploits the channel structure in each dimension (mode) simultaneously. The one-bit CS-based CSI feedback method in \cite{8902107} enables a simple and cost-effective construction of the quantizer. In \cite{9126231}, the channel matrices are compressed using random projection at the UE, where deep learning (DL) is employed to recover the original CSI. The classical CS principle leverages the sparse structure of the signal in reconstruction, while the channel matrix is usually not perfectly sparse but nearly sparse. The modeling error in the CS-based CSI feedback methods leads to an inevitable CSI reconstruction error.

\par
The state-of-the-art technology for CSI feedback exploits the powerful DL technology that is capable of learning nonlinear feature representations of the CSI. A DL architecture, named CsiNet \cite{8322184}, uses an encoder to learn a transformation from CSI to codewords in the UE and an inverse transformation to reconstruct the CSI from these codewords in the BS. It is improved into CsiNet+ \cite{8972904} that exploits the sparsity characteristics of CSI in the angular-delay domain, and has a larger receptive field, more refine blocks and a deeper network architecture. In \cite{9296555}, the fully-convolutional neural network architecture, called DeepCMC, with residual layers at the decoder, incorporates quantization and entropy coding blocks into its design. DeepCMC is trained to minimize a weighted rate-distortion cost, which enables a trade-off between the CSI quality and its feedback overhead. Beyond the use of convolution layers, there are other effective designs for CSI feedback. Long short term memory (LSTM) is applied in \cite{8482358} to extract the temporal correlation of CSI for enhancing recovery performance. Recurrent neural network (RNN) in both the encoder and decoder is used in \cite{8543184,8951228} to learn the temporal correlation. CQNet proposed in \cite{9090892} jointly tackles CSI compression, codeword quantization, and recovery under the bandwidth constraint. Deep transfer learning is applied in \cite{9442844} to solve the problem of high training cost associated with downlink CSI feedback networks. The DL-based CSI feedback framework in \cite{9279228} maximizes the gain in beamforming performance rather than the accuracy of the feedback. It should be noted that these DL-based methods follow the idea of source compression and assume a perfect wireless feedback channel in which no error occurs. Considering the case of an imperfect feedback channel, the CSI feedback framework in \cite{xu2022deep} exploits the idea of deep joint source-channel coding (DJSCC), which is superior to the separate source-channel coding scheme and is more robust to channel degradation. Interested readers are referred to \cite{guo2022overview} for more related work on DL-based CSI feedback.

\par
In most existing DL-based feedback methods \cite{8322184,8972904,xu2022deep,8482358,8543184,8951228,9495802,9585309,9419066,9497358,9373670}, e.g., CsiNet, CsiNet+ and DJSCC, the DL models on the BS side and the UE side are highly coupled and need to be trained jointly, leading to several drawbacks. First, the two-sided DL models require collaborations between different network vendors and UE vendors in the training and inference phases, which raises various issues to be considered in order to achieve consensus, e.g., model maintenance and responsibility. Second, it is unlikely for UEs to store many DL models due to the limited storage space. For the two-sided DL models, delivering the UE side model from the BS to UEs consumes extra spectrum resource. Third, complex DL models are not suitable for low-cost UEs with limited device memory and computing capability. It is desired to conduct simple operations at the UEs, while the BS can handle computationally intensive tasks.

\par
The idea of using neural networks solely at the BS for CSI recovery is introduced in CS-CsiNet \cite{8322184}, which learns to recover CSI from random linear measurements. Subsequently, different one-sided DL architecture are proposed \cite{9126231,wang2022deep}. However, existing one-sided DL-based methods can only compress the CSI matrix with a fixed compression ratio (CR). The BS has to train and store several DL models to realize multi-rate CSI compression. The multi-rate CSI feedback approach in \cite{8972904} simultaneously trains several DL models for different CRs, where the models for different CRs share a portion of the common parameters. Although a reduction of 40\% storage space is reported in \cite{8972904}, the DL model is suitable only for CRs that have been considered in training and further reduction of storage space is still very challenging.

\par
In this paper, we propose a novel one-sided one-for-all framework for DL-based CSI feedback. Specifically, the CSI is compressed simply via a small number of linear projections at the UE, and is recovered at the BS in an iterative manner, which involves the use of DL. The one-for-all property means that only one DL model is needed to deal with arbitrary CRs. This method comes from the popular plug-and-play priors (PPP), which was first introduced for image reconstruction \cite{6737048} and then achieved success in many signal processing tasks. The main idea is to unroll the CSI reconstruction problem by the variable splitting technique and replace the prior associated sub-problem by any off-the-shelf CSI denoising methods. Instead of using handcrafted regularizers for the wireless channel responses, we treat the denoising as a black-box, and capture the angular-delay characteristics existing in the CSI data. Furthermore, the denoise solver can be learned by exploiting DL, which has a large capability in feature abstraction and gives rise to promising performance. We would like to emphasize that the proposed one-sided one-for-all framework for DL-based CSI feedback is different from those CS-based methods \cite{6214417,6966062,9126231}, except for the linear projection operation in UEs. We employ a flexible deep PPP \cite{9413947,9454311,9325040} that iteratively applies DL as a denoising operator to exploit the characteristics of CSI in the BS. Deep PPP does not require knowledge of the UE process in the training phase and is therefore agnostic to the UE process, leading to the one-for-all property of the proposed method, while the learning-based one-for-one CS networks in \cite{8322184,9126231,wang2022deep} are subject to some specified projection matrix and CR, and cannot be reused for other projection matrices and CRs.

\begin{figure*}[t]
\centering
\includegraphics[scale=0.4]{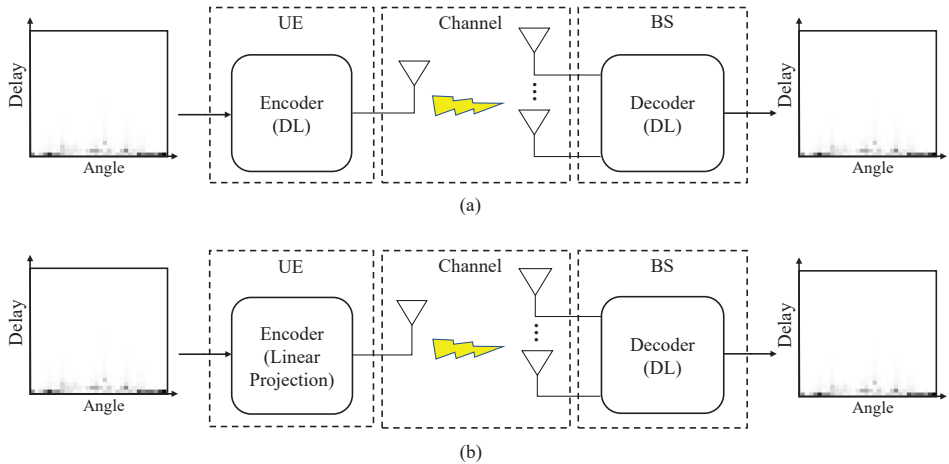}
\caption{DL-based CSI feedback frameworks. (a) The two-sided framework, e.g., CsiNet~\cite{8322184}, CsiNet+~\cite{8972904} and DJSCC-Net~\cite{xu2022deep}; (b) The one-sided framework.}
\label{system}
\end{figure*}

\par
The proposed CSI feedback framework has several advantages. First, the DL model at the BS is totally decoupled with operations at the UEs, and is trained without knowing what linear operations are performed to compress the CSI at the UEs. Thus, only the network vendors take the responsibility of maintaining the DL model, and the trained DL model can be used for various CSI projection designs of different UE vendors. Second, the proposed framework requires only the delivery of the projection design from the UE to the BS, which causes very light overhead, e.g., only transmitting the random variable seed that generates the random projection matrix. Third, the proposed framework is suitable for UEs with limited device memories and computation power, which conduct linear projections instead of complex DL models.


\par
The rest of the paper is organized as follows. Section \uppercase\expandafter{\romannumeral2} introduces the background and system model of CSI feedback. Section \uppercase\expandafter{\romannumeral3} provides the proposed CSI recovery method, which exploits deep PPP, and presents details of the network architecture and training process. The experimental results are presented in Section \uppercase\expandafter{\romannumeral4}, followed by the conclusions in Section \uppercase\expandafter{\romannumeral5}.


\section{System Model}
We consider a single cell massive MIMO frequency orthogonal division multiplexing (OFDM) system operating in FDD mode. There is a uniform linear array (ULA)\footnote{The proposed method can also be applied to other types of antennas. Here we consider the ULA model for simplicity and fair comparison with other methods in the literature.} with $N_t$ antennas equipped in the BS and each UE has a single antenna. Different UEs work at distinct frequency bands. The downlink CSI in the spatial-frequency domain is represented by $\mathbf{\tilde{H}}\in \mathbb{C}^{N_s \times N_t}$, where $N_s$ denotes the number of subcarriers assigned to the UE. In the FDD system, the UE should send $\mathbf{\tilde{H}}$ back to the BS for precoding. In this paper, we assume that perfect CSI has been acquired through pilot-based channel estimation and focus on channel feedback.

\par
To reduce feedback overhead, we could represent the channel matrices in the angular-delay domain using a 2D discrete Fourier transform (DFT), which can be described as following
\begin{equation}\label{eq:model}
\mathbf{H} =\mathbf{F}_s\mathbf{\tilde{H}}\mathbf{F}_t,
\end{equation}
where $\mathbf{F}_s$ and $\mathbf{F}_t$ are $N_s \times N_s$ and $N_t \times N_t$ DFT matrices, respectively. The channel matrix in the angular-delay domain contains only a small fraction of large coefficients, and the other are close to zero. Furthermore, as the time delay between multipath arrivals lies within a limited period, almost all large coefficients are in the first a few rows of $\mathbf{H}$ in the delay domain. Therefore, in~\cite{8322184,8972904,9126231}, the first a few rows of $\mathbf{H}$ are fed back to the BS and the remaining rows are ignored. It is found in \cite{xu2022deep} that some useful information about the CSI is discarded in the truncation process at the UE. In this paper, we develop a MIMO CSI feedback approach that can be applied to the above cases to further improve their performance. We use $\mathbf{H}$ to denote the truncated channel matrix by abuse of notation.

\subsection{The Two-Sided DL-Based CSI Feedback Framework}
Most existing DL-based methods for CSI feedback utilize the two-sided autoencoder framework \cite{8322184,8972904,xu2022deep,8482358,8543184,8951228,9495802,9585309,9419066,9497358,9373670}, which is a type of DL technique used to learn efficient data encoding. The encoder model at the UE maps the CSI matrix to a low-dimensional compressed space, while the decoder model at the BS maps the received feedback information to the original dimension to construct an approximation of the CSI. It naturally overcomes the limits of CS-based approaches that enforce channel sparsity. These existing works mainly focus on the design of the encoder/decoder DL modules with various considerations, e.g., utilizing more powerful DL building blocks to achieve better performance.

\par
The two-sided DL-based CSI feedback framework is shown in Fig. \ref{system} (a), where the UE-sided model and the BS-sided model could be jointly trained, e.g., at a single side/entity. This leads to a high overhead cost in delivering the DL model from one side to the other side. Separate training on the BS side and on the UE side requires collaborations, which also results in some sacrifice of communication resources. In addition, the use of DL comes at a considerable cost of computational complexity, which significantly hinders its applications in low-cost UEs, e.g., IoT devices.

\subsection{The One-Sided DL-Based CSI Feedback Framework}
The one-sided CSI feedback framework is illustrated in Fig. \ref{system} (b), where only the BS-side requires the deployment of DL models. At the UE, the CSI is simply compressed via a few linear projections, which significantly alleviates the burden of computation at the UE. The compression process can be expressed as
\begin{equation}\label{eq:cs}
\mathbf{y} =\mathbf{A}\mathbf{h}+\mathbf{n},
\end{equation}
where $\mathbf{h}\in \mathbb{R}^{N}$ is composed of the real and imaginary parts of the vectored $\mathbf{H}\in \mathbb{C}^{N_s\times N_t}$, $N=2N_s N_t$, and $\mathbf{y}\in \mathbb{R}^{M}$ and $\mathbf{A}\in \mathbb{R}^{M\times N}$ denote the compressed CSI vector and the linear projection matrix, respectively, and $\mathbf{n}\in \mathbb{R}^{M}$ is the additive noise caused by quantization to the compressed signal. The dimension of the feedback CSI, $M$, is much less than the dimension of the channel $N$, and therefore the CR $\frac{M}{N}$ is much less than one. 

\par
The linear projection matrix $\mathbf{A}$, which is also called a sensing matrix or a measurement matrix in CS, could be generated randomly. Hence by knowing the random variable seed that is used to generate $\mathbf{A}$ in the UE, the BS could produce the linear projection matrix $\mathbf{A}$ with a negligible amount of communication overhead. The transmission of the random variable seed, e.g., a few bits, is much lighter than the burden of model delivery in the two-sided feedback framework. Although we employ the same process in UE as CS-based methods~\cite{6214417,6966062,8902107,9126231} and one-sided learning-based methods \cite{8322184,9126231,wang2022deep}, the proposed CSI-PPPNet for CSI reconstruction is very different from these existing methods. First, the proposed method exploits a DL model learned directly from CSI training data, which avoids the strict sparse assumption, i.e., the prerequisite of CS-based methods \cite{6214417,6966062,9126231}. Second, the proposed method offers generalization against various linear projection operations in the UEs, as the model training is agnostic to the compression process. This means that the trained DL model in the BS can be applied to arbitrary projection matrices $\mathbf{A}$, while one-sided learning-based methods \cite{8322184,9126231,wang2022deep} are subject to some specified projection matrix and CR, and cannot be reused for other projection matrices and CRs.

\section{CSI-PPPNet}
In this section, we present the proposed CSI recovery method, namely CSI-PPPNet, which involves a DL model that is agnostic to the process in the UEs.

\subsection{CSI Recovery With the PPP}
The recovery of the CSI from compressed feedback in (\ref{eq:cs}) could be boiled down to a Maximum A Posteriori (MAP) estimation problem
\begin{equation}\label{eq:MAP}
\mathbf{\hat{h}}=\arg\max\limits_{\mathbf{h}}\log p(\mathbf{y}|\mathbf{h})+ \log p(\mathbf{h}),
\end{equation}
where $p(\mathbf{y}|\mathbf{h})$ and $p(\mathbf{h})$ denote the likelihood of feedback $\mathbf{y}$ and the prior of the channel matrix $\mathbf{H}$, respectively. The MAP estimation problem can be reformulated as
\begin{equation}\label{eq:CSI_recovery}
\begin{split}
\min_{\mathbf{h}} \quad &\|\mathbf{y}-\mathbf{Ah}\|_2^2 +\lambda J(\mathbf{h}),
\end{split}
\end{equation}
where $J(\cdot)$ presents the regularizer incorporating the prior information to enforce the desired delay-angle property of the channel responses, and $\lambda$ is a positive penalty parameter to control the impact of the regularizer.

\par
There are a few challenges in directly dealing with the CSI recovery problem in (\ref{eq:CSI_recovery}). First, it is not a trivial task to handcraft an effective regularizer to capture the property of the channel responses. Furthermore, it is not desired to use complex regularizers due to the difficulty of algorithm development. In this paper, we provide a flexible and extendable framework to address this problem, in which any off-the-shelf DL-based denoiser can be exploited.

\par
We first introduce an auxiliary variable $\mathbf{z}$ and the CSI recovery problem in (\ref{eq:CSI_recovery}) can be rewritten as
\begin{equation}\label{eq:CSI_recovery_auxiliary}
\begin{split}
\min_{\mathbf{h},\mathbf{z}} \quad &\|\mathbf{y}-\mathbf{Az}\|_2^2 +\lambda J(\mathbf{h}),\\
\text{s.t.} \quad & \mathbf{z} = \mathbf{h}.
\end{split}
\end{equation}
Various algorithms have been developed to solve this kind of problem. Here, we consider solving it by minimizing the Lagrangian function
\begin{equation}\label{eq:CSI_recovery_Lagrangian}
\begin{split}
L_{\rho}(\mathbf{h},\mathbf{z})= &\|\mathbf{y}-\mathbf{Az}\|_2^2 +\lambda J(\mathbf{h})+\rho\|\mathbf{h}-\mathbf{z}\|_2^2,
\end{split}
\end{equation}
where $\rho$ is a positive penalty parameter. By gradually increasing $\rho$, the solution of (\ref{eq:CSI_recovery_Lagrangian}) tends to be close to the solution of (\ref{eq:CSI_recovery_auxiliary}). This optimization problem can be addressed by iteratively solving the following subproblems for $\mathbf{z}$ and $\mathbf{h}$ while keeping the rest of the variables fixed,
\begin{equation}\label{eq:HQS_Z}
\mathbf{z}^{t+1} := \arg\min_{\mathbf{z}} \ \|\mathbf{y}-\mathbf{Az}\|_2^2 +\rho\|\mathbf{h}^t-\mathbf{z}\|_2^2,
\end{equation}
\begin{equation}\label{eq:HQS_H}
\mathbf{h}^{t+1} := \arg\min_{\mathbf{h}}\ \lambda J(\mathbf{h})+\rho\|\mathbf{h}-\mathbf{z}^t\|_2^2.
\end{equation}
The subproblem in (\ref{eq:HQS_Z}) can be solved by the least-square method, which gives
\begin{equation}\label{eq:HQS_Z_LS}
\mathbf{z}^{t+1} = \left(\mathbf{A}^T\mathbf{A}+\rho\mathbf{I}\right)^{-1}\left(\mathbf{A}^T\mathbf{y}+\rho\mathbf{h}^t\right).
\end{equation}
The subproblem in (\ref{eq:HQS_H}) can be rewritten as
\begin{equation}\label{eq:HQS_H2}
\mathbf{h}^{t+1} := \arg\min_{\mathbf{h}}\  J(\mathbf{h})+\frac{\rho}{\lambda}\|\mathbf{h}-\mathbf{z}^t\|_2^2.
\end{equation}

\begin{figure*}[t]
\centering
\includegraphics[scale=0.50]{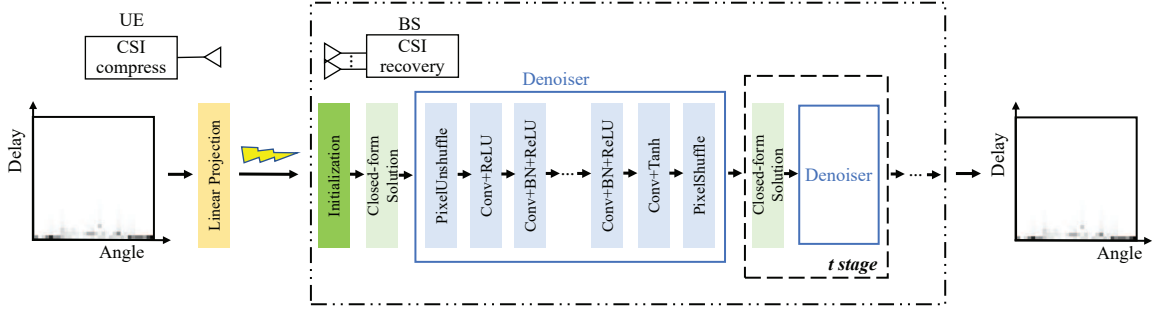}
\caption{The proposed one-sided one-for-all framework with CSI-PPPNet.}
\label{CSIPPP}
\end{figure*}

\par
From a Bayesian perspective, the problem in (\ref{eq:HQS_H2}) can be regarded as denoising $\mathbf{z}^{t}$ in the presence of additive Gaussian noise with a standard deviation $\sqrt{\frac{\lambda}{2\rho}}$ according to the MAP estimation. Instead of handcrafting a regularizer $J$, we could consider (\ref{eq:HQS_H2}) as a denoising problem,
\begin{equation}\label{eq:HQS_H3}
\mathbf{h}^{t+1}= \mathcal{D}\left(\mathbf{z}^{t},\sqrt{\frac{\lambda}{2\rho}}\right).
\end{equation}
The denoising process, $\mathcal{D}$, captures the delay-angle characteristics of the channel responses. With the various choices of denoising methods, the CSI recovery method is flexible and extendable. The procedure of the proposed CSI recovery with the PPP is summarized in Algorithm 1.

\par
Solving the subproblem in (\ref{eq:HQS_H}) can be seen as a proximal operation, which is replaced by a denoiser in the PPP method. It is interesting to investigate under which conditions a denoiser can actually be a proximal map, and to analyze the convergence of the algorithm. In \cite{7542195}, the Moreau theorem is used to prove the convergence of the PPP method, while the analysis is limited to non-expansive denoisers. Imposing non-expansiveness on the denoiser is very likely to negatively affect its denoising performance \cite{ryu2019plug}. As indicated in~\cite{hurault2022proximal}, the imposing of the convergence result is extended to deep denoisers that are not non-expansive. In the experimental study, we provide empirical convergence to show the working mechanism of the proposed deep PPP for CSI recovery.

\begin{algorithm}[t]
\caption{The proposed CSI Recovery with the PPP} \label{alg1}
\begin{algorithmic}
\STATE \hspace*{-5mm} Input: The feedback $\mathbf{y}$, the linear projection matrix $\mathbf{A}$, the regularization parameter $\lambda$, the penalty factor $\rho^1$ and the scaling factor $\alpha$.
\STATE \hspace*{-5mm} Output: The recovered CSI vector, $\mathbf{\hat{h}}$;
\STATE \hspace*{-5mm} Step 1: Initialize the CSI variable $\mathbf{h}^0$;
\STATE \hspace*{-5mm} Step 2: Compute $\mathbf{z}^{t+1}$ using (\ref{eq:HQS_Z_LS});
\STATE \hspace*{-5mm} Step 3: Compute $\mathbf{h}^{t+1}$ using (\ref{eq:HQS_H3});
\STATE \hspace*{-5mm} Step 4: Update $\rho^{t+1}= \alpha\rho^{t}$;
\STATE \hspace*{-5mm} Step 5: Let $t=t+1$, and go to step 2 if some halting condition is not satisfied.
\end{algorithmic}
\end{algorithm}

\subsection{Network Architecture and Training}
The proposed one-sided framework with CSI-PPPNet is shown in Fig. \ref{CSIPPP}. On the UE side, we reshape the real and imaginary parts of the normalized CSI matrix $\mathbf{H}$ into a vector with real value $\mathbf{h}\in\mathbb{R}^{N}$. Then we apply linear projections to the CSI vector, $\mathbf{h}$, to obtain the feedback vector $\mathbf{y}\in\mathbb{R}^{M}$. After the BS obtains the feedback vector, $\mathbf{y}$, the CSI-PPPNet is conducted to map it back to the CSI matrix $\mathbf{H}$. As described in Algorithm \ref{alg1} and shown in Fig. \ref{CSIPPP}, the proposed method CSI-PPPNet consists mainly of two parts, i.e., the closed-form solution in (\ref{eq:HQS_Z_LS}) and the denoiser, which are performed in an alternating way.

\par
At the beginning, we initialize $\mathbf{h}$ in the BS with a rough CSI estimate. Considering the sparsity of the CSI vector $\mathbf{h}$, we initialize it through a simple non-iterative process. In particular, we apply the least-square estimation on $\mathbf{h}$ according to the support corresponding to some of the largest elements in $\mathbf{A}^T\mathbf{y}$. The initialization is described in Algorithm \ref{alg2}. One could also initialize $\mathbf{h}$ with a vector with all zeros, while more iterations are usually needed to achieve good recovery results.

\par
The design of the denoiser module is critical. On the one hand, as CSI-PPPNet is an iterative algorithm, the denoiser module needs to be computationally light, but without sacrificing much denoising performance. On the other hand, as the recovery error of the CSI would gradually decrease in iterations, the denoiser needs to have the ability to adapt to varying noise variance. With the development of DL technologies, convolutional neural network (CNN) based image denoisers, such as DnCNN~\cite{7839189}, IRCNN~\cite{8099783} and FDDNet~\cite{8365806}, have shown excellent performance in terms of denoising effectiveness and denoising efficiency. If using DnCNN, we need to train models for each noise level separately to ensure the effectiveness of CSI-PPPNet, which requires a lot of storage space and training efforts. IRCNN consists of 25 independent 7-layer denoisers, each of which is trained at noise levels within a small range, so it can handle a range of noise levels. However, this training strategy also lacks flexibility to handle arbitrary noise levels. Unlike DnCNN and IRCNN, the FDDNet inputs are four subimages generated by downsampling the original noisy image, and a noise level map generated by an adjustable parameter $\sigma$. The output of the FDDNet is four denoised sub-images, which form the whole denoised image via reversible upsampling. This unique design allows the network to be guided by the noise level, which makes the network highly adaptable to varying noise levels. In addition, denoising on sub-images improves the network receptive field without increasing the network complexity.


\par
The denoiser used in CSI-PPPNet is shown in Fig. \ref{CSIPPP}. We reshape the noisy CSI vector $\mathbf{z}^t\in\mathbb{R}^{N}$ into a tensor $\mathbf{Z}^t\in\mathbb{R}^{N_s \times N_t \times 2}$ and then input it into the network. Inspired by FDDNet, the first layer of the denoiser is the pixel unshuffle layer \cite{8365806}, which divides the noisy CSI matrix into four sub-tensors of size $\frac{N_s}{2} \times \frac{N_t}{2} \times 2$. Then we generate a uniform noise level map of size $\frac{N_s}{2} \times \frac{N_t}{2}$ with all elements being $\sigma$, and concatenate the noise level map with the four sub-tensors into a larger tensor $\mathbf{\hat{Z}}^t\in\mathbb{R}^{\frac{N_s}{2}\times \frac{N_t}{2} \times 9}$. Afterwards, the tensor $\mathbf{\hat{Z}}^t$ is fed into a series of $3 \times 3$ convolutional layers, each consisting of a specific combination of four operations: convolution (Conv), rectified linear unit (ReLU), batch normalization (BN) and TanHyperbolic (Tanh). Specifically, the first convolutional layer, the middle layer, and the last convolutional layer adopt the combination of ``Conv+ReLU", ``Conv+BN+ReLU" and ``Conv+Tanh", respectively. The convolutional layers use the zero padding to ensure that the size of the feature map remains unchanged after each convolution. The benefit of using BN in the middle layer is two-fold. It would make the network converge faster, and the output of the convolutional layer more stable. In the last convolutional layer, the tanh function is used to scale values in the range of $[-1, 1]$. Lastly, the pixel shuffle layer is used as a reverse operation of the pixel unshuffle to generate the denoised CSI. Considering the balance between complexity and performance, we set 8 convolutional layers with 48 convolution kernels in each iteration.

\par
Notably, the DL denoiser can be used for different CRs. For the training stage, we normalize the original clean CSI $\mathbf{H}$ and then generate the noisy CSI dataset by adding random Gaussian noise from a variety of noise levels. Each training data sample includes the original clean CSI $\mathbf{H}$, the noisy CSI $\mathbf{\check{H}}$ and the noise level $\sigma$. The denoised channel can be expressed as,
\begin{equation}\label{eq:denoiser}
\mathbf{\hat{H}} = Denoiser(\mathbf{\check{H}}, \sigma; \Theta),
\end{equation}
where $\Theta$ represents the parameter set of the denoiser, and $\sigma$ denotes the noise level of the CSI matrix. To learn model parameters from noisy CSI dataset, we use the loss function given by
\begin{equation}\label{eq:loss}
L(\Theta) = \frac{1}{T} \sum_{j=1}^{T} \frac{\left \|  \mathbf{H} _{j} - Denoiser( \mathbf{\check{H}} _{j}, \sigma_{j} ;\Theta) \right \|_{F}^{2}}{\left \| \mathbf{H} _{j}  \right \|_{F}^{2}} ,
\end{equation}
where $T$ is the total number of samples in the training set, the subscript denotes the $j$th sample in the training set and $\|\cdot\|_F$ denotes the Frobenius norm.

\begin{algorithm}[t]
\caption{Initialization.} \label{alg2}
\label{alg:Init}
\begin{algorithmic}
\STATE \hspace*{-5mm} Input: The feedback $\mathbf{y}$, the linear projection matrix $\mathbf{A}$, the sparsity $K$;
\STATE \hspace*{-5mm} Output: Initialized CSI vector, $\mathbf{h}^0$;
\STATE \hspace*{-5mm} Step 1: Find the set ${\Omega}$ that includes indices corresponding to the top $K$ largest elements in $\mathbf{A}^T\mathbf{y}$;
\STATE \hspace*{-5mm} Step 2: Conduct the least squares estimation, $\mathbf{h}_{{\Omega}} = \arg\min_{\mathbf{h}_{{\Omega}}}\|\mathbf{y} - \mathbf{A}_{{\Omega}}\mathbf{h}_{{\Omega}}\|_2^2=(\mathbf{A}_{\Omega}^T\mathbf{A}_{\Omega})^{-1}\mathbf{A}_{\Omega}^T\mathbf{y}$;
\STATE \hspace*{-5mm} Step 3: Obtain $\mathbf{h}^0$ with nonzero elements according to $\mathbf{h}_{{\Omega}}$.
\end{algorithmic}
\end{algorithm}

\section{Experimental Results}
In this section, we evaluate the performance of the proposed CSI-PPPNet, i.e., one-sided framework for massive MIMO CSI feedback. Downlink CSI is generated by QuaDRiGa~\cite{AuaDRiGa} according to the 3rd Generation Partnership Project (3GPP) TR 38.901~\cite{3GPP38901}. For the evaluation, we consider two scenarios, i.e., open indoor and urban macro (UMa). The center frequency, inter-site distance, the height of BS and the height of UE are 5.2GHz, 20m, 3m and 1.5m for the indoor scenario, and 2.4GHz, 400m, 25m and 1.5m for the UMa scenario, respectively. There are $N_s = 256$ subcarriers in the OFDM system, and following the setting in~\cite{8322184,8972904,8482358,8543184,8951228,9495802,9585309,9419066,9497358,9373670}, the first $N_s=32$ rows that contain most nonzero values are truncated for compression. The BS is equipped with $N_t = 32$ antennas, while the UE has one antenna. Both the antennas in the BS and in the UE are omnidirectional. Other specific parameters are shown in Table \ref{table of data para}.

\begin{table}
\renewcommand\arraystretch{1.3}
\caption{Simulation Settings for Indoor and UMa.}
\centering
\begin{tabular}{l|l}
\hline
\multicolumn{1}{c|}{\textbf{Parameters}} & \multicolumn{1}{c}{\textbf{Values}}  \\
\hline
Scenarios                                & Indoor, UMa                          \\
\hline
BS antenna configurations                & 32 elements, omnidirectional         \\
\hline
UE mobility                              & 3km/h                                \\
\hline
Bandwidth                                & 10MHz                                \\
\hline
UT antenna configurations                & 1 element, omnidirectional           \\
\hline
Center Frequency                         & 5.2GHz for Indoor, 2.4GHz for UMa    \\
\hline
Carrier Number                           & 256                                  \\
\hline
Inter-site Distance                      & 20m for Indoor, 400m for UMa         \\
\hline
BS height                                & 3m for Indoor, 25m for UMa           \\
\hline
UE height                                & 1.5m                                 \\
\hline
\end{tabular}
\label{table of data para}
\end{table}

\par
The generated CSI after normalization forms the training dataset for the denoising network and the evaluation dataset for the proposed CSI-PPPNet and other competitors. In the training stage of the denoising network, we generate the noisy CSI by adding random Gaussian noise with the signal-to-noise ratio (SNR) drawn uniformly in the range of $[0,40]dB$. Each sample includes the original clean CSI, the noisy CSI and the noise level. The training and validation datasets contain 100,000 and 30,000 samples, respectively. We set the batch size to 128. Adam optimizer is first initialized with a learning rate of $10^{-4}$. When the loss does not decrease in 20 epochs, the learning rate will be reduced by half. The lower bound of the learning rate is set to $10^{-7}$. Then we generate 20,000 clean CSI samples to evaluate the CSI recovery accuracy and the achievable downlink rate for the proposed CSI feedback method and competitors. To generate a linear projection matrix $\mathbf{A}$, we apply the singular value decomposition to an $N \times N$ random matrix and use the first $M$ singular vectors to form the $M$ rows of $\mathbf{A}$. The proposed CSI recovery with the PPP stops after 10 iterations. The regularization parameter, $\lambda$, the penalty factor, $\rho$, and the scaling factor, $\alpha$, are fine-tuned for each CSI CR. The difference between the recovered channel matrix, $\mathbf{\hat{H}}$, and the original channel matrix, $\mathbf{H}$, is quantified by the normalized mean-squared error (NMSE), which is expressed as
\begin{equation}\label{eq:NMSE}
\text{NMSE}= \mathbb{E}\left(\frac{\|\mathbf{\hat{H}}-\mathbf{H}\|_F^2}{\|\mathbf{H}\|_F^2}\right).
\end{equation}

%

\begin{figure*}[t]
\centering
\subfigure{\includegraphics[width=0.75\linewidth]{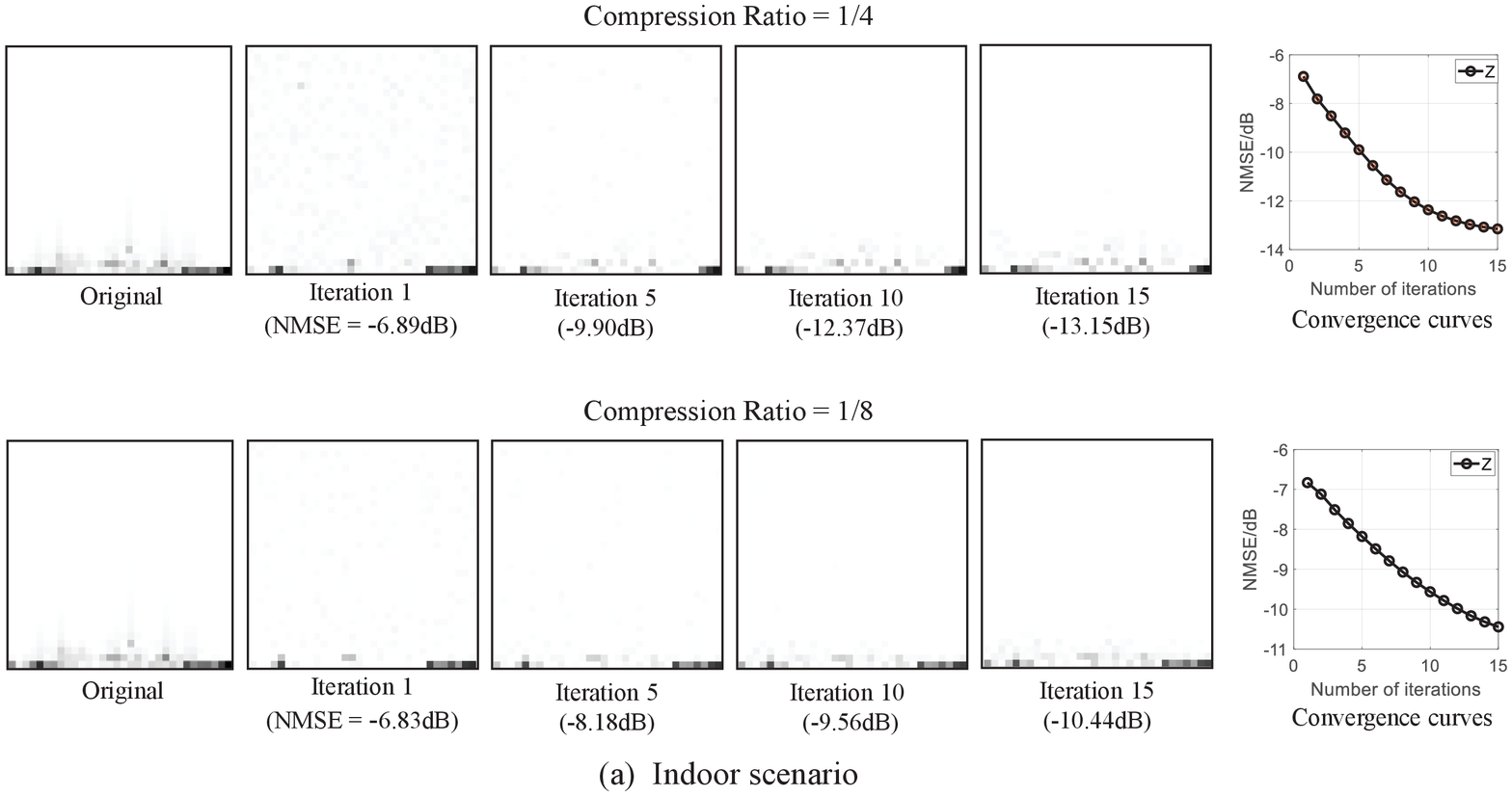}
\label{convergence_indoor}}
\subfigure{\includegraphics[width=0.76\linewidth]{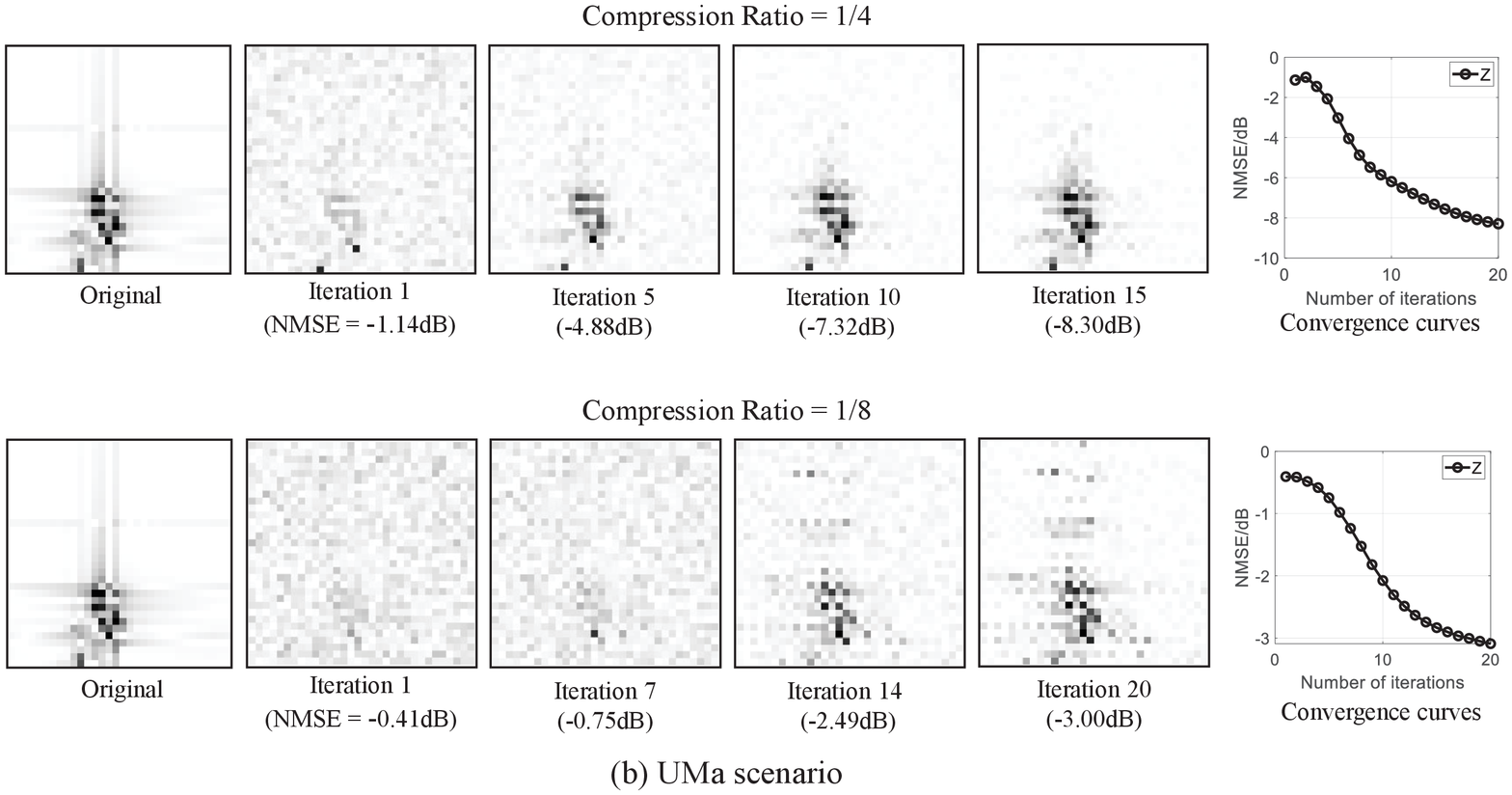}
\label{convergence_uma}}
\caption{Illustration of the convergence of the proposed CSI reconstruction method with CSI-PPPNet for the indoor and the UMa scenarios. Left: Visual and NMSE results of the reconstructed CSI matrix at different iterations; Right: Convergence curve of NMSE (y-axis) with respect to the number of iterations (x-axis).} \label{convergence}
\end{figure*}
In addition to the NMSE that measures the CSI reconstruction error, the cosine similarity (CoS) is another popular CSI feedback accuracy measure used in the literature~\cite{8322184,8972904,9126231}. As the feedback CSI serves as a beamforming vector, the CoS measures the quality of the beamforming vector, which is given as
\begin{equation}\label{eq:cosine}
\text{CoS}= \mathbb{E}\left(\frac{1}{N_s}\sum_{i=1}^{N_s}\frac{|\mathbf{\hat{h}}^H_i\mathbf{h}_i|}{\|\mathbf{\hat{h}}_i\|_2\|\mathbf{h}_i\|_2}\right),
\end{equation}
where $\mathbf{\hat{h}}_i$ and $\mathbf{h}_i$ denote the recovered channel vector and the original channel vector of the $i$th subcarrier, respectively.

\subsection{Convergence}
The proposed CSI-PPPNet implicitly exploits the DL-based denoising operator as the regularizer in (\ref{eq:CSI_recovery}), which cannot generally be expressed as a proximal mapping. Therefore, the proposed algorithm does not seek the minimization of an explicit objective function. Recent theoretical work has analyzed the convergence of the PPP algorithms. Sreehari et al. present sufficient conditions that ensure convergence of the PPP approach \cite{7542195}. Specifically, it requires the denoising operator to be a proximal mapping, which holds if it is nonexpansive and its subgradient is a symmetric matrix. Subsequent research provides the convergence guarantee for various specific conditions, e.g., bounded denoisers~\cite{7744574}, continuous denoisers \cite{8237460} and Gaussian mixture model (GMM) based denoisers \cite{8168194}. Generally speaking, the convergence guarantee of the PPP algorithms requires some strong assumptions on the denoiser, which usually do not hold in DL based denoisors, but extensive experimental results show that DL based denoisors work well in the PPP framework \cite{9454311,9325040,ryu2019plug}.

\par
To better understand the working mechanism of the proposed CSI-PPPNet for CSI recovery, we illustrate the empirical convergence in Fig. \ref{convergence}. Subfigures on the left show the visual results and NMSE results of the closed-form solution, $\mathbf{z}^t$, at different iterations for the indoor and the UMa scenarios. As observed, the recovered CSI matrices in the first iteration contain large errors compared to the original CSI matrices. After the results are passed through the DL-based denoiser and the closed-form solution was updated several times, the unwanted artifacts are removed and the CSI structures are retained, resulting in a decrease of the NMSE. The subfigures on the right show the NMSE converges quickly. 


\subsection{CSI Feedback Accuracy Without Quantization}
\label{CSI Feedback Accuracy without quantization}
To evaluate the accuracy of the CSI feedback, we compare the one-sided one-for-all CSI-PPPNet method with the classical CsiNet, which is a two-sided autoencoder-based method that applies DL models in both the UE and the BS. As the UE uses linear projection, we also compare the proposed method with other one-sided frameworks, such as CS-CsiNet in \cite{8322184}, ReNet in \cite{9126231}, and TVAL3~ in \cite{li2009user}. TVAL3 is a CS-based algorithm, which has been shown to be superior to several other CS-based algorithms for CSI reconstruction in \cite{9126231}. 

\begin{figure*}[t]
\centering
\includegraphics[scale=0.45]{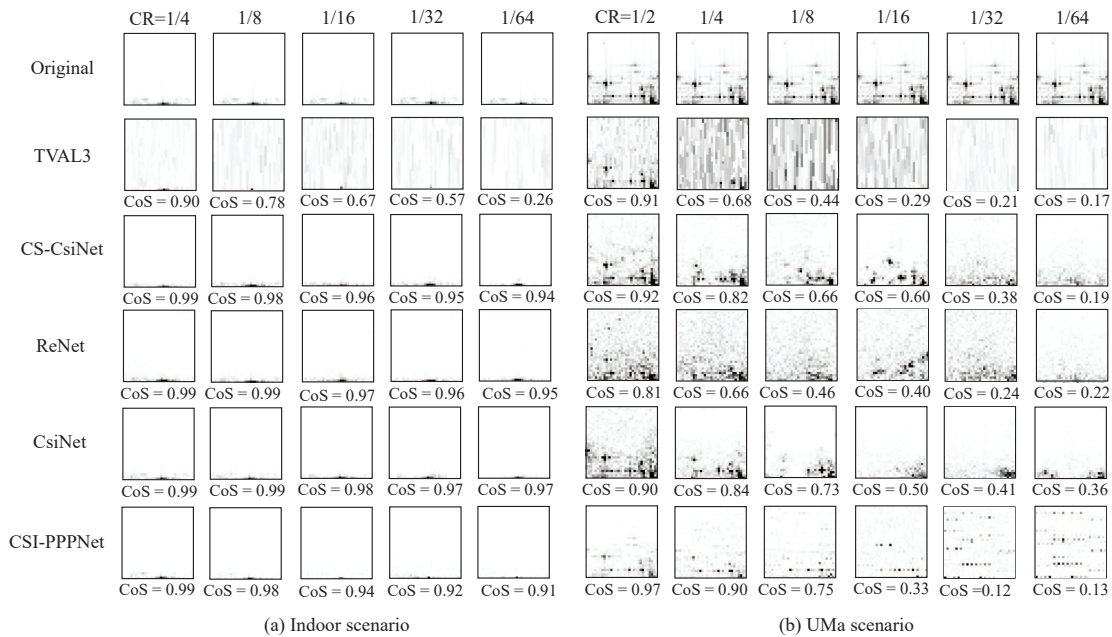}
\caption{Visual results comparison of different CSI feedback methods in the indoor scenario and the UMa scenario.}
\label{visual}
\end{figure*}
\par
Table \ref{acc_indoor} and Table \ref{acc_uma} show the CSI feedback accuracy for the indoor scenario and the UMa scenario, respectively. The proposed method outperforms all other methods at the CR of $1/2$ and $1/4$ in the UMa scenario. For TVAL3, based on CS, the proposed method is better at all CRs in the indoor scenario. Compared to CS-CsiNet, the performance of the proposed method is better with CRs from $1/16$ to $1/64$ in the indoor scenario. It is not surprising that the two-sided autoencoder-based method achieves better CSI feedback accuracy in most cases, as the DL-based encoder at the UE could extract more compact features than the simple linear projection operation, especially when the CR is low. Therefore, the one-sided framework can be regarded as the result of trading-off the CSI feedback accuracy with the cost in both the encoder complexity and the model training/storage/delivery complexity. Furthermore, for existing methods in the literature, DL-based decoder needs to be jointly trained with the encoder and/or linear projections, while the proposed method has the advantage of the one-for-all property that further reduces the model training/storage/delivery complexity, i.e., a DL model trained once for CSI denoising can be repurposed for CSI recovery tasks with arbitrary linear projections for all UEs and CRs. Interestingly, learning-based methods, that is, CsiNet and CSI-PPPNet, achieve more accurate CoS than the CS-based TVAL3 algorithm, when their NMSE performance is close. For example, in the indoor scenario, CSI-PPPNet with CR $1/32$ and the TVAL3 algorithm with CR $1/8$ achieve similar NMSE, while their CoS are $0.79$ and $0.71$, respectively. This phenomenon suggests that learning-based methods would be more promising if considering the construction of a precoded vector using the feedback CSI.

\begin{table}
\setlength{\abovecaptionskip}{-0.05cm} 
\setlength{\belowcaptionskip}{-0.2cm}
\renewcommand\arraystretch{1.2}
\centering
\caption{Comparison of the CSI Feedback Accuracy without Quantization for the Indoor Scenario}
\renewcommand\arraystretch{1}
\begin{tabular}{cl|cc}
\hline
\textbf{CR}   & \textbf{Method} & \textbf{NMSE}   & \textbf{CoS}                \\
\hline
\multirow{5}{*}{1/4}
                      & TVAL3                                         & -10.42 & 0.90               \\
                      & CS-CsiNet                                     & -16.18 & 0.99               \\
                      & ReNet                                         & -14.60 & 0.99              \\
                      & CsiNet                                        & -19.16 & 0.99               \\
                      & CSI-PPPNet                                    & -16.32 & 0.99               \\
\hline
\multirow{5}{*}{1/8}
                      & TVAL3                                         & -3.49  & 0.71               \\
                      & CS-CsiNet                                     & -14.09 & 0.98              \\
                      & ReNet                                         & -14.60 & 0.99              \\
                      & CsiNet                                        & -18.12 & 0.99               \\
                      & CSI-PPPNet                                    & -10.08 & 0.95               \\
\hline
\multirow{5}{*}{1/16}
                      & TVAL3                                         & -1.45  & 0.55               \\
                      & CS-CsiNet                                     & -1.62 & 0.57
                       \\
                      & ReNet                                         & -11.14 & 0.96              \\
                      & CsiNet                                        & -15.88 & 0.99               \\
                      & CSI-PPPNet                                    & -6.75  & 0.90               \\
\hline
\multirow{5}{*}{1/32}
                      & TVAL3                                         & -0.46  & 0.35               \\
                      & CS-CsiNet                                     & -0.75 & 0.43
                       \\
                      & ReNet                                         & -9.22 & 0.94              \\
                      & CsiNet                                        & -12.78 & 0.98               \\
                      & CSI-PPPNet                                    & -3.43  & 0.79               \\
\hline
\multirow{5}{*}{1/64}
                      & TVAL3                                         & -0.12  & 0.21               \\
                      & CS-CsiNet                                     & -0.27 & 0.29
                       \\
                      & ReNet                                         & -7.82 & 0.92              \\
                      & CsiNet                                        & -9.48  & 0.94               \\
                      & CSI-PPPNet                                    & -0.56  & 0.62               \\
\hline
\end{tabular}
\label{acc_indoor}
\end{table}

\begin{table}
\setlength{\abovecaptionskip}{-0.05cm} 
\setlength{\belowcaptionskip}{-0.2cm}
\renewcommand\arraystretch{1.2}
\centering
\caption{Comparison of the CSI Feedback Accuracy without Quantization for the UMa Scenario}
\renewcommand\arraystretch{1}
\begin{tabular}{cl|cc}
\hline
\textbf{CR}   & \textbf{Method} & \textbf{NMSE}   & \textbf{CoS}                \\
\hline
\multirow{5}{*}{1/2}  & TVAL3                                         & -10.05 & 0.93            \\
                      & CS-CsiNet                                     & -6.58  & 0.88            \\
                      & ReNet                                         & -4.12  & 0.79            \\
                      & CsiNet                                        & -6.39  & 0.88            \\
                      & CSI-PPPNet                                    & -11.05 & 0.96            \\
\hline
\multirow{5}{*}{1/4}  & TVAL3                                         & -3.90  & 0.68            \\
                      & CS-CsiNet                                     & -4.59  & 0.81            \\
                      & ReNet                                         & -2.64  & 0.68            \\
                      & CsiNet                                        & -5.13  & 0.84            \\
                      & CSI-PPPNet                                    & -6.47  & 0.89            \\
\hline
\multirow{5}{*}{1/8}  & TVAL3                                         & -1.77  & 0.50            \\
                      & CS-CsiNet                                     & -2.70  & 0.69            \\
                      & ReNet                                         & -1.61  & 0.56            \\
                      & CsiNet                                        & -3.79  & 0.76            \\
                      & CSI-PPPNet                                    & -2.61  & 0.73            \\
\hline
\multirow{5}{*}{1/16} & TVAL3                                         & -0.59  & 0.30            \\
                      & CS-CsiNet                                     & -1.61  & 0.57            \\
                      & ReNet                                         & -1.04  & 0.47            \\
                      & CsiNet                                        & -2.17  & 0.63            \\
                      & CSI-PPPNet                                    & -0.83   & 0.48           \\
\hline
\multirow{5}{*}{1/32} & TVAL3                                         & -0.17  & 0.20            \\
                      & CS-CsiNet                                     & -0.45  & 0.36            \\
                      & ReNet                                         & -0.49  & 0.36            \\
                      & CsiNet                                        & -1.36  & 0.53            \\
                      & CSI-PPPNet                                    & 1.84   & 0.22           \\
\hline
\multirow{5}{*}{1/64} & TVAL3                                         & -0.05  & 0.17            \\
                      & CS-CsiNet                                     & -0.24  & 0.28            \\
                      & ReNet                                         & -0.18  & 0.26            \\
                      & CsiNet                                        & -1.05  & 0.48            \\
                      & CSI-PPPNet                                    & 2.51   & 0.094           \\
\hline
\end{tabular}
\label{acc_uma}
\end{table}

\par
Fig. \ref{visual} shows visual results comparison of different CSI feedback methods in the indoor scenario and the UMa scenario. In the figure, the CSI example of the indoor scenario is more sparse than the example of the UMa scenario. Therefore, for all different methods with the same CR, the CoS performance of the UMa scenario is lower than the performance of the indoor scenario. The TVAL3 method introduces some noticeable false structures, especially when the CR is low. Such odd structures are highly reduced by learning-based methods, i.e., CsiNet, CS-CsiNet, ReNet and CSI-PPPNet.

\begin{figure}[!tb]
\centering
\includegraphics[scale=0.4]{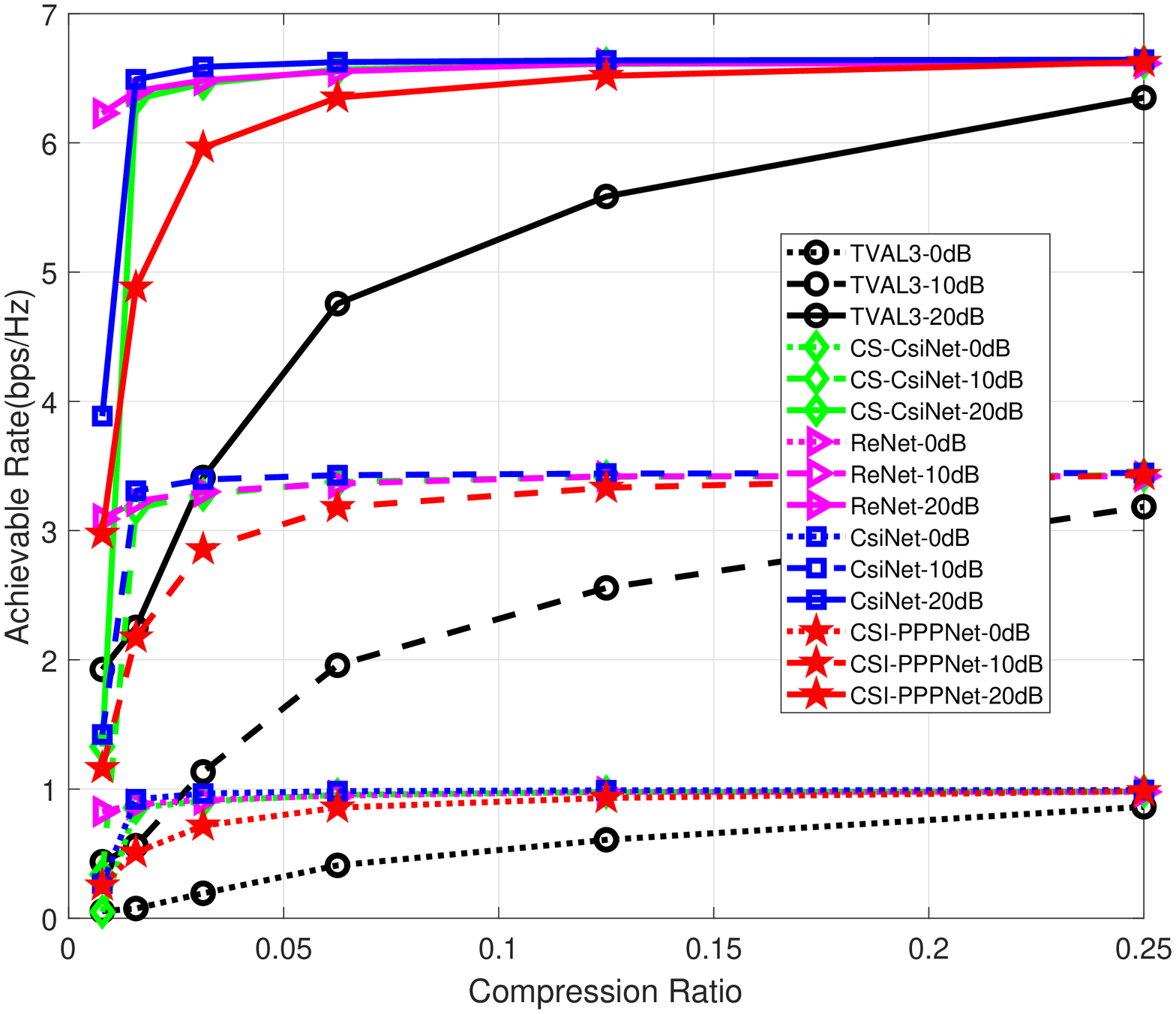}
\caption{Comparison of achievable downlink rate with CSI feedback in the indoor scenario.}
\label{achievable_rate_indoor}
\end{figure}
\par

\begin{figure}[!tb]
\centering
\includegraphics[scale=0.4]{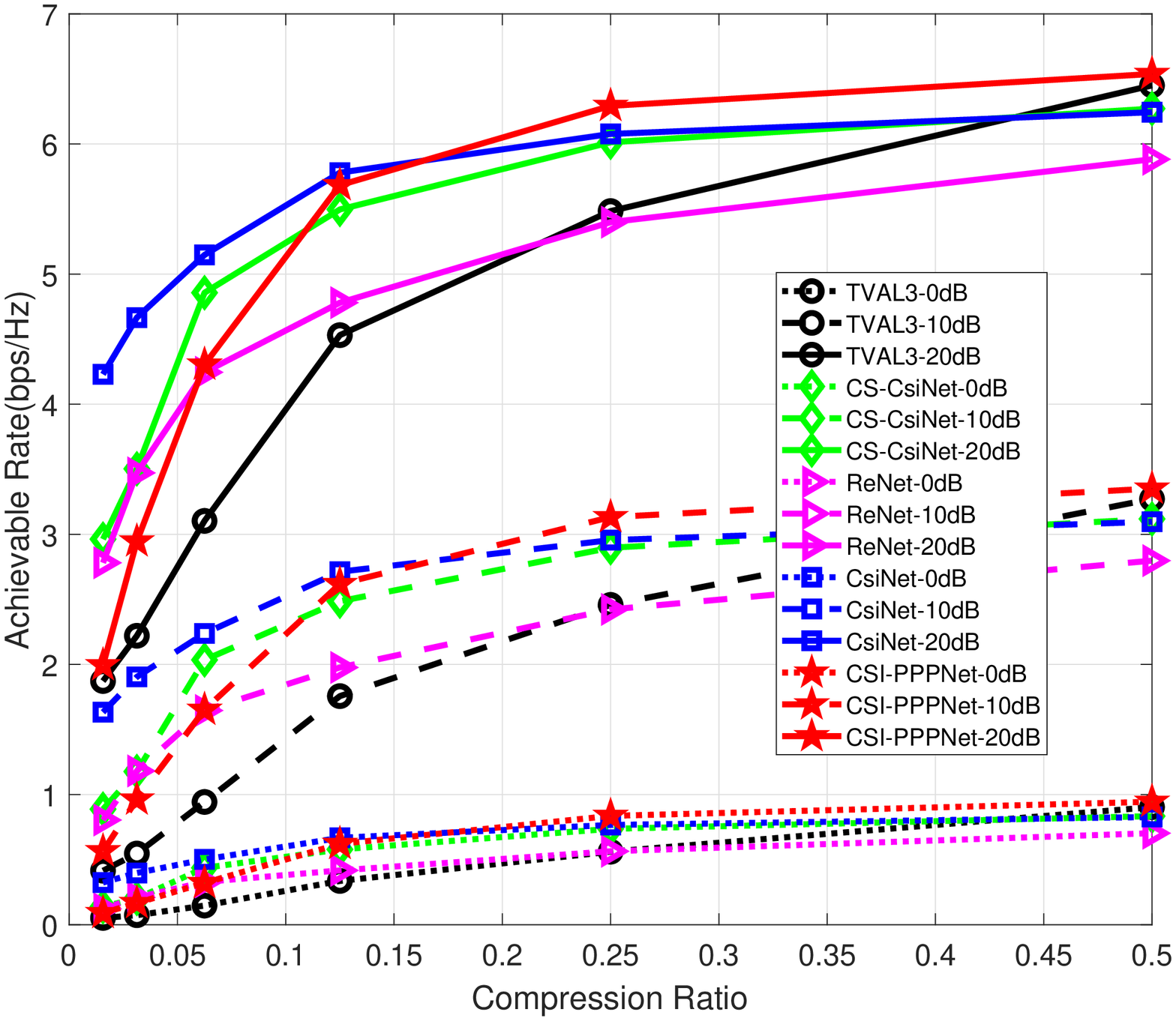}
\caption{Comparison of achievable downlink rate with CSI feedback in the UMa scenario.}
\label{achievable_rate_uma}
\end{figure}

\subsection{Achievable Downlink Rate With CSI Feedback}
Furthermore, we consider the achievable downlink transmission rate with CSI feedback in the evaluation. With the information of CSI feedback, the BS could use the precoding technology in massive MIMO systems to increase throughput. Here, we employ the matched filter (MF) precoding, as it requires low processing complexity. The MF precoding is also known as maximum ratio transmission and conjugate beamforming. The precoding vector for the $i$th subcarrier is simply the conjugate transpose of the downlink CSI vector,
\begin{equation}\label{eq:precoding}
\mathbf{w}_i=\frac{\mathbf{\hat{h}}_i^H}{\|\mathbf{\hat{h}}_i\|_2}.
\end{equation}
Then the average achievable downlink rate is
\begin{equation}\label{eq:achievable_rate}
R = \mathbb{E}\left(\frac{1}{N_s}\sum_{i=1}^{N_s}\log_2 \left(1+\text{SNR}_{DL}|\mathbf{w}_i\mathbf{h}_i|^2  \right) \right),
\end{equation}
where $\text{SNR}_{DL}$ denotes the downlink SNR. With accurate CSI feedback, we have a larger product between $\mathbf{w}_i$ and $\mathbf{h}_i$ and thus obtain a high achievable downlink rate.

\par
Fig. \ref{achievable_rate_indoor} and Fig. \ref{achievable_rate_uma} show the achievable downlink rate with CSI feedback in the indoor and the UMa scenarios, respectively. Downlink SNRs of $0$ dB, $10$ dB, and $20$ dB are considered in the evaluation. The proposed CSI-PPPNet achieves the highest achievable downlink rate in the UMa scenario for high CRs. Although CsiNet performs the best for the indoor scenario, the gap between CsiNet and CSI-PPPNet is very small when the CR is greater than 0.1.

\subsection{CSI Feedback Accuracy With Quantization}
Since CsiNet does not consider quantization in CSI compression, in this experiment, CsiNet+ \cite{guo2020convolutional} which considers the quantization operation is used for comparison. We adopt the non-uniform quantization in [45] in all the compared methods, where the value of $\mu$ is 200 and the numbers of quantization bits are chosen as 3, 4, 5 and 6.

\begin{table}[!htb]\small
	\setlength{\abovecaptionskip}{-0.05cm} 
	\setlength{\belowcaptionskip}{-0.2cm}
	\renewcommand{\arraystretch}{1.2} 
	\centering
	\caption{Comparison of the CSI Feedback Accuracy (NMSE/CoS) with Quantization for the Indoor Scenario}
    \renewcommand\arraystretch{1}
	\begin{tabular}{c|c|ccccc}
		\hline
		\textbf{Method} & \diagbox{\textbf{B}}{\textbf{CR}} & 1/4 & 1/8 & 1/16 & 1/32 & 1/64 \\
		\hline
		\multirow{5}{*}{TVAL3} & 3 & -7.58/0.79 & -3.18/0.63 & -1.24/0.49 & -0.46/0.34 & -0.13/0.21 \\
		& 4 & -9.30/0.86 & -3.84/0.68 & -1.44/0.54 & -0.47/0.35 & -0.13/0.22 \\
		& 5 & -10.13/0.89 & -4.09/0.70 & -1.47/0.55 & -0.47/0.35 & -0.14/0.22 \\
		& 6 & -10.46/0.90 & -4.12/0.71 & -1.48/0.55 & -0.47/0.35 & -0.14/0.22 \\
		& no quantization & -10.63/0.91 & -4.14/0.71 & -1.46/0.56 & -0.47/0.35 & -0.14/0.22 \\
		\hline
            \multirow{5}{*}{CS-CsiNet} 
            & 3      & 4.56/0.49   & 9.76/0.32   & 14.17/0.21  & 12.15/0.19 & 11.00/0.16 \\
            & 4      & -1.87/0.77  & 4.24/0.53   & 9.15/0.34   & 9.55/0.26  & 6.61/0.23  \\
            & 5      & -7.00/0.91     & -1.64/0.76  & 3.49/0.56   & 3.17/0.55  & 0.25/0.58  \\
            & 6      & -11.58/0.97 & -7.06/0.91  & -3.40/0.82  & -5.01/0.85 & -4.20/0.82 \\
            & no quantization & -16.18/0.99 & -14.09/0.98  & -10.96/0.96 & -7.64/0.92 & -7.08/0.90 \\
		\hline
            \multirow{5}{*}{ReNet} 
            & 3      & 7.46/0.27   & 10.23/0.22 & 12.95/0.20  & 11.71/0.18 & 12.01/0.16  \\
            & 4      & 3.98/0.48   & 7.73/0.35  & 9.84/0.29   & 7.78/0.25  & 5.52/0.22  \\
            & 5      & -1.09/0.73  & 3.15/0.56  & 5.73/0.45   & 2.06/0.54  & -0.29/0.55  \\
            & 6      & -6.28/0.89  & -2.42/0.79 & -1.22/0.74  & -4.09/0.82 & -4.05/0.81  \\
            & non-quantization & -14.60/0.99 & -14.60/0.99 & -11.14/0.96 & -9.22/0.94  & -7.82/0.92 \\
		\hline
		\multirow{5}{*}{CsiNet+} & 3 & -15.01/0.98 & -13.19/0.98 & -12.20/0.97 & -9.63/0.95 & -8.49/0.93 \\
		& 4 & -17.53/0.99 & -16.21/0.99 & -14.87/0.98 & -11.07/0.96 & -9.02/0.94 \\
		& 5 & -18.81/0.99 & -18.25/0.99 & -16.58/0.99 & -11.96/0.97 & -9.24/0.94 \\
		& 6 & -19.28/0.99 & -19.11/0.99 & -17.25/0.99 & -12.30/0.97 & -9.30/0.94 \\
		& no quantization & -19.31/0.99 & -19.21/0.99 & -17.45/0.99 & -12.26/0.97 & -9.25/0.94 \\
		\hline
		\multirow{5}{*}{CSI-PPPNet} & 3 & -16.20/0.99 & -12.03/0.97 & -6.51/0.89 & -3.11/0.78 & -0.19/0.56 \\
		& 4 & -16.27/0.99 & -12.10/0.97 & -6.69/0.90 & -3.34/0.79 & -0.42/0.58 \\
		& 5 & -16.30/0.99 & -12.11/0.97 & -6.76/0.90 & -3.40/0.79 & -0.51/0.59 \\
		& 6 & -16.31/0.99 & -12.12/0.97 & -6.77/0.90 & -3.42/0.79 & -0.51/0.59 \\
		& no quantization & -16.32/0.99 & -12.13/0.97 & -6.75/0.90 & -3.43/0.79 & -0.56/0.62 \\
		\hline
	\end{tabular}%
	\label{table:03:07}%
\end{table}%

\begin{table}[!htb]\small
	\setlength{\abovecaptionskip}{-0.05cm} 
	\setlength{\belowcaptionskip}{-0.2cm}
	\renewcommand{\arraystretch}{1.2} 
	\centering
	\caption{Comparison of the CSI Feedback Accuracy (NMSE/CoS) with Quantization for the UMa Scenario}
    \renewcommand\arraystretch{1}
	\begin{tabular}{c|c|cccccc}
		\hline
		\textbf{Method} & \diagbox{\textbf{B}}{\textbf{CR}} & 1/2   & 1/4   & 1/8   & 1/16 & 1/32 & 1/64\\
		\hline
		\multirow{5}[2]{*}{TVAL3} & 3     & -9.26/0.91 & -3.65/0.63 & -1.51/0.44 & -0.58/0.29 & -0.18/0.202  & -0.06/0.17 \\
		& 4     & -9.28/0.91 & -4.25/0.69 & -1.73/0.48 & -0.59/0.30 & -0.17/0.20 &  -0.04/0.17\\
		& 5     & -9.29/0.91 & -4.40/0.71 & -1.76/0.49 & -0.60/0.30 &  -0.17/0.20 & -0.05/0.17\\
		& 6     & -9.29/0.91 & -4.45/0.71 & -1.77/0.49 & -0.60/0.30 &  -0.17/0.20 & -0.05/0.17\\
		& no quantization   & -10.05/0.93 & -4.49/0.71 & -1.77/0.49 & -0.60/0.30  & -0.17/0.20      & -0.05/0.17 \\                 
		\hline
        \multirow{5}[2]{*}{CS-CsiNet} & 3     & 10.07/0.32 & 11.00/0.27    & 11.18/0.22 & 7.57/0.21  & 4.64/0.20  & 3.15/0.18  \\
        & 4     & 2.15/0.57  & 2.5/0.50   & 2.14/0.42  & 1.03/0.35  & 0.38/0.26  & 0.28/0.21  \\
        & 5     & -4.12/0.80 & -3.21/0.75 & -1.59/0.62 & -1.03/0.50 & -0.32/0.33 & -0.11/0.25 \\
        & 6     & -6.02/0.87 & -4.30/0.80  & -2.48/0.68 & -1.46/0.55 & -0.42/0.35 & -0.21/0.27 \\
        & no quantization & -6.58/0.88 & -4.59/0.81 & -2.70/0.69 & -1.61/0.57 & -0.45/0.36 & -0.24/0.28 \\
		\hline
        \multirow{5}[2]{*}{ReNet} & 3 & 9.69/0.22 & 7.04/0.20 & 5.16/0.16 & 2.84/0.17 & 1.49/0.17 & 1.81/0.17 \\
        & 4     & 4.29/0.39 & 2.50/0.36 & 0.92/0.31 & 0.59/0.26 & 0.36/0.23 &  0.20/0.20\\
		& 5     & -0.70/0.61 & -0.81/0.54 & -0.63/0.45 & -0.54/0.40 & -0.26/0.31 & -0.07/0.23\\
		& 6     & -3.08/0.73 & -2.10/0.64 & -1.35/0.53 & -0.88/0.45 & -0.41/0.34 & -0.15/0.25\\
		& no quantization   & -4.12/0.79 & -2.64/0.68 & -1.61/0.56 & -1.04/0.47  & -0.49/0.36 & -0.18/0.26 \\ 
		\hline
		\multirow{5}[2]{*}{CsiNet+} & 3     & -7.60/0.91 & -5.03/0.82 & -2.00/0.63 & -0.95/0.50 & -1.11/0.48 & -0.66/0.40\\
		& 4     & -8.35/0.92 & -5.47/0.84 & -2.15/0.64 & -1.03/0.51 & -1.16/0.49 & -0.69/0.41\\
		& 5     & -8.59/0.93 & -5.60/0.84 & -2.19/0.65 & -1.05/0.51 &  -1.17/0.49 & -0.70/0.41\\
		& 6     & -8.66/0.93 & -5.63/0.84 & -2.21/0.65 & -1.05/0.51 & -1.18/0.50 & -0.70/0.41\\
		& no quantization   & -8.64/0.93 & -5.64/0.84 & -2.21/0.65 & -1.05/0.51 &-1.16/0.49 & -0.67/0.40\\

		\hline
		\multirow{5}[2]{*}{CSI-PPPNet} & 3     & -11.03/0.96 & -6.39/0.88 & -2.41/0.71 & 0.70/0.40 & 2.76/0.04 & 2.84/0.03\\
		& 4     & -11.04/0.96 & -6.44/0.88 & -2.55/0.72 & 0.55/0.42 & 1.96/0.20 & 2.84/0.03\\
		& 5     & -11.04/0.96 & -6.46/0.88 & -2.60/0.72 & 0.51/0.43 & 1.87/0.22 & 2.84/0.03 \\
		& 6     & -11.04/0.96 & -6.46/0.88 & -2.60/0.72 & 0.49/0.43 & 1.85/0.22 & 2.84/0.03 \\
		& no quantization   & -11.05/0.96 & -6.47/0.89 & -2.61/0.73 & -0.83/0.48 & 1.84/0.22 & 2.51/0.09\\
		\hline
	\end{tabular}%
	\label{table:03:08}%
\end{table}%

\par
Table \ref{table:03:07} and \ref{table:03:08} show the CSI feedback accuracy of different methods in the indoor and the UMa scenarios, respectively. The proposed CSI-PPPNet outperforms all other competitors at CRs from $1/2$ to $1/8$ and all quantization levels in the UMa scenario. Under the Indoor scenario, the performance of CsiNet+ is optimal, which is consistent with the experimental results without quantization, as the DL-based encoder at UE can extract more compact features than a simple linear projection, especially when the CR is relatively low or the CSI structure is simpler. 

\par
The total feedback overhead is related to both the compression rate and the quantization level. For example, in the UMa scenario, 3 bits quantization with 1/2 CR has the same feedback cost as 6 bit quantization with 1/4 CR. By comparing the NMSE of the three methods in these two cases, it can be found that the CSI reconstruction performance is better under a higher CR with a lower quantization level. Thus, instead of increasing the number of quantized bits, it is better to increase the CR in the UMa scenario. However, this phenomenon is not observed in the indoor scenario.

\par
Finally, compared to all the competitors, it is interesting to note that CSI-PPPNet is not sensitive to the change of the quantization level, which indicates that CSI-PPPNet has a better adaptability to the noise brought by quantization. We think the use of the denoising network in the CSI-PPPNet helps in dealing with different levels of quantization noise adaptively.

\subsection{Simulation Results on the COST 2100 Data Set}
In this subsection, we use the public CSI dataset, which is in the indoor scenario with the COST 2100 channel model and used in existing works for performance evaluation. We compare the accuracy of the CSI feedback of our proposed method with that of other methods reported in \cite{9126231}. The results are summarized in Table \ref{cost2100}.

\par
The CSI-PPPNet has the best performance among all compared methods with the CR 1/4. It outperforms the two-sided model CsiNet at the CR 1/8. However, at lower CRs from 1/16 to 1/64, its performance is inferior to CS-CsiNet and ReNet. Note that CSI-PPPNet only uses one model trained and stored at the BS, while other methods require different models for different CRs. The number of model parameters for these methods is discussed in the next subsection.

\begin{table}[!ht]
\centering
\caption{Comparison of different methods under the COST2100 Indoor}
\begin{tabular}{l|l|c|l|l|l}
\hline
\textbf{
\begin{tabular}[c]{@{}l@{}}
\diagbox{Method} {CR}  
\end{tabular}} & \multicolumn{1}{c|}{1/4} & \multicolumn{1}{c|}{1/8} & \multicolumn{1}{c|}{1/16} & \multicolumn{1}{c|}{1/32} & \multicolumn{1}{c}{1/64} \\ \hline
TVAL3 & -14.87/0.97    & - / -    & -2.61/0.66     & -0.27/0.33      & -0.63/0.11       \\
CS-CsiNet & -11.82/0.95   & - / -  & -6.09/0.86    & -4.67/0.80  & -2.46/0.66            \\
CsiNet    & -17.36/0.97  & -12.07/0.96      & -8.65/0.91 & -6.24/0.86  & -5.84/0.84   \\
ReNet     & -17.68/0.99  & -12.88/0.97      & -9.09/0.93 & -6.34/0.87  & -3.44/0.76   \\
CSI-PPPNet    & -18.42/0.99     & -12.17/0.97   & -5.16/0.84    & -2.24/0.64      & -0.65/0.37      \\ \hline                 
\end{tabular}
\label{cost2100}
\end{table}

\subsection{Model Storage}
Now we compare the number of model parameters stored in the UE and in the BS to support the CSI feedback with five different CRs, including $\frac{1}{4}$, $\frac{1}{8}$, $\frac{1}{16}$, $\frac{1}{32}$ and $\frac{1}{64}$. The conventional iterative alternative direction algorithm, TVAL3, does not need any model training, so it has no model storage overhead. The number of parameters is given in Table \ref{storage}. The numbers of parameters are the same in both the indoor scenario and the UMa scenario, so we do not differentiate the scenarios when presenting the results. CsiNet exploits the autoencoder with an encoder model and a decoder model. The deployment of the DL model imposes an additional burden of storage and computation on the UE. As shown in Table \ref{storage}, more than 4 million parameters are needed in CsiNet to support the five CRs. In comparison, we exploit the one-sided DL framework, where the UE compresses the CSI with a random projection matrix generated by some random variable seed. Thus, the UE only needs to store and feeds back one extra parameter, i.e., the random variable seed, to the BS. 
\par
In addition, CSI-PPPNet has an appealing one-for-all property. This means only one DL-based denoiser model needs to be trained and stored at the BS, and is used for arbitrary CRs. However, all the other learning-based methods need to train different models for distinct CRs, and thus their model storage overhead increases with the number of supported CRs. To support the five CRs in the experiment, the proposed method requires $175.2$K parameters in total, which is only $4.3\%$, $1.2\%$ and $8.5\%$ of the number of parameters in CsiNet, ReNet and CS-CsiNet, respectively.

\begin{table}
\renewcommand\arraystretch{1.8}
\centering
\caption{Comparison of Model Storage Overhead}
\begin{tabular}{l|c|c|c}
\hline
\textbf{Method} & \textbf{UE params} & \textbf{BS params} & \textbf{Total params}  \\
\hline
CS-CsiNet-$\left\{\frac{1}{4},\frac{1}{8},\frac{1}{16},\frac{1}{32},\frac{1}{64}\right\}$ & 1 & 2062K & 2062K     \\
\hline
ReNet-$\left\{\frac{1}{4},\frac{1}{8},\frac{1}{16},\frac{1}{32},\frac{1}{64}\right\}$ & 1 & 14909K & 14909K     \\
\hline
CsiNet-$\left\{\frac{1}{4},\frac{1}{8},\frac{1}{16},\frac{1}{32},\frac{1}{64}\right\}$ & 2033.4K & 2062K & 4095.4K     \\
\hline
CSI-PPPNet              & 1         & 175.2K   & 175.2K       \\
\hline
\end{tabular}
\label{storage}
\end{table}

\section{Conclusion}
In this work, a novel one-sided one-for-all DL framework is proposed for massive MIMO CSI feedback. CSI is compressed through linear projections at the UE, which could be applied at UEs with limited device memory and computation power. Then the CSI is recovered at the BS using DL. We explore the strategy of using the PPP to decouple the DL model training and the UEs' compression procedure, which avoids joint model training and model delivery in traditional two-sided DL models for massive MIMO CSI feedback. The proposed method has an appealing one-for-all property, that is, a DL-based denoiser model for arbitrary CSI CRs, which significantly reduces the number of models for training and storage in the BS. Our experiments show the advantages of the proposed CSI feedback method.


%
%

\ifCLASSOPTIONcaptionsoff
  \newpage
\fi



%

\bibliographystyle{IEEEtran}
\bibliography{IEEEabrv,bib_paper}

\begin{thebibliography}{10}
\providecommand{\url}[1]{#1}
\csname url@samestyle\endcsname
\providecommand{\newblock}{\relax}
\providecommand{\bibinfo}[2]{#2}
\providecommand{\BIBentrySTDinterwordspacing}{\spaceskip=0pt\relax}
\providecommand{\BIBentryALTinterwordstretchfactor}{4}
\providecommand{\BIBentryALTinterwordspacing}{\spaceskip=\fontdimen2\font plus
\BIBentryALTinterwordstretchfactor\fontdimen3\font minus
  \fontdimen4\font\relax}
\providecommand{\BIBforeignlanguage}[2]{{%
\expandafter\ifx\csname l@#1\endcsname\relax
\typeout{** WARNING: IEEEtran.bst: No hyphenation pattern has been}%
\typeout{** loaded for the language `#1'. Using the pattern for}%
\typeout{** the default language instead.}%
\else
\language=\csname l@#1\endcsname
\fi
#2}}
\providecommand{\BIBdecl}{\relax}
\BIBdecl

\bibitem{6736761}
E.~G. Larsson, O.~Edfors, F.~Tufvesson, and T.~L. Marzetta, ``Massive {MIMO}
  for next generation wireless systems,'' \emph{IEEE Communications Magazine},
  vol.~52, no.~2, pp. 186--195, 2014.

\bibitem{8861014}
L.~Sanguinetti, E.~Bjornson, and J.~Hoydis, ``Toward massive {MIMO} 2.0:
  Understanding spatial correlation, interference suppression, and pilot
  contamination,'' \emph{IEEE Transactions on Communications}, vol.~68, no.~1,
  pp. 232--257, 2020.

\bibitem{3GPP5}
RP-182863, ``{MIMO} evolution for downlink and uplink,'' \emph{{3rd Generation
  Partnership Project (3GPP), RAN Meeting 94e,}}, 2021.

\bibitem{3GPP4}
38.214, ``{NR} physical layer procedures for data (release 17),'' \emph{3rd
  Generation Partnership Project (3GPP), Technical Specification,}, 2022.

\bibitem{6214417}
P.-H. Kuo, H.~T. Kung, and P.-A. Ting, ``Compressive sensing based channel
  feedback protocols for spatially-correlated massive antenna arrays,'' in
  \emph{2012 IEEE Wireless Communications and Networking Conference (WCNC)},
  2012, pp. 492--497.

\bibitem{6966062}
P.~Cheng and Z.~Chen, ``Multidimensional compressive sensing based analog {CSI}
  feedback for massive {MIMO-OFDM} systems,'' in \emph{2014 IEEE 80th Vehicular
  Technology Conference (VTC2014-Fall)}, 2014, pp. 1--6.

\bibitem{8902107}
C.~Qing, Q.~Yang, B.~Cai, B.~Pan, and J.~Wang, ``Superimposed coding-based
  {CSI} feedback using 1-bit compressed sensing,'' \emph{IEEE Communications
  Letters}, vol.~24, no.~1, pp. 193--197, 2020.

\bibitem{9126231}
P.~Liang, J.~Fan, W.~Shen, Z.~Qin, and G.~Y. Li, ``Deep learning and
  compressive sensing-based {CSI} feedback in {FDD} massive {MIMO} systems,''
  \emph{IEEE Transactions on Vehicular Technology}, vol.~69, no.~8, pp.
  9217--9222, 2020.

\bibitem{8322184}
C.-K. Wen, W.-T. Shih, and S.~Jin, ``Deep learning for massive {MIMO CSI}
  feedback,'' \emph{IEEE Wireless Communications Letters}, vol.~7, no.~5, pp.
  748--751, 2018.

\bibitem{8972904}
J.~Guo, C.-K. Wen, S.~Jin, and G.~Y. Li, ``Convolutional neural network-based
  multiple-rate compressive sensing for massive {MIMO CSI} feedback: Design,
  simulation, and analysis,'' \emph{IEEE Transactions on Wireless
  Communications}, vol.~19, no.~4, pp. 2827--2840, 2020.

\bibitem{9296555}
M.~B. Mashhadi, Q.~Yang, and G{\"u}nd{\"u}z, ``Distributed deep convolutional
  compression for massive {MIMO CSI} feedback,'' \emph{IEEE Transactions on
  Wireless Communications}, vol.~20, no.~4, pp. 2621--2633, 2021.

\bibitem{8482358}
T.~Wang, C.-K. Wen, S.~Jin, and G.~Y. Li, ``Deep learning-based {CSI} feedback
  approach for time-varying massive {MIMO} channels,'' \emph{IEEE Wireless
  Communications Letters}, vol.~8, no.~2, pp. 416--419, 2019.

\bibitem{8543184}
C.~Lu, W.~Xu, H.~Shen, J.~Zhu, and K.~Wang, ``{MIMO} channel information
  feedback using deep recurrent network,'' \emph{IEEE Communications Letters},
  vol.~23, no.~1, pp. 188--191, 2019.

\bibitem{8951228}
X.~Li and H.~Wu, ``Spatio-temporal representation with deep neural recurrent
  network in {MIMO CSI} feedback,'' \emph{IEEE Wireless Communications
  Letters}, vol.~9, no.~5, pp. 653--657, 2020.

\bibitem{9090892}
Z.~Liu, L.~Zhang, and Z.~Ding, ``An efficient deep learning framework for low
  rate massive {MIMO CSI} reporting,'' \emph{IEEE Transactions on
  Communications}, vol.~68, no.~8, pp. 4761--4772, 2020.

\bibitem{9442844}
J.~Zeng, J.~Sun, G.~Gui, B.~Adebisi, T.~Ohtsuki, H.~Gacanin, and H.~Sari,
  ``Downlink {CSI} feedback algorithm with deep transfer learning for {FDD}
  massive {MIMO} systems,'' \emph{IEEE Transactions on Cognitive Communications
  and Networking}, vol.~7, no.~4, pp. 1253--1265, 2021.

\bibitem{9279228}
J.~Guo, C.-K. Wen, and S.~Jin, ``Deep learning-based {CSI} feedback for
  beamforming in single- and multi-cell massive {MIMO} systems,'' \emph{IEEE
  Journal on Selected Areas in Communications}, vol.~39, no.~7, pp. 1872--1884,
  2021.

\bibitem{xu2022deep}
J.~Xu, B.~Ai, N.~Wang, and W.~Chen, ``Deep joint source-channel coding for
  {CSI} feedback: An end-to-end approach,'' \emph{IEEE Journal on Selected
  Areas in Communications}, vol.~41, no.~1, pp. 260--273, 2023.

\bibitem{guo2022overview}
J.~Guo, C.-K. Wen, S.~Jin, and G.~Y. Li, ``Overview of deep learning-based
  {CSI} feedback in massive {MIMO} systems,'' \emph{arXiv preprint
  arXiv:2206.14383}, 2022.

\bibitem{9495802}
Z.~Hu, J.~Guo, G.~Liu, H.~Zheng, and J.~Xue, ``{MRFNet}: A deep learning-based
  {CSI} feedback approach of massive {MIMO} systems,'' \emph{IEEE
  Communications Letters}, vol.~25, no.~10, pp. 3310--3314, 2021.

\bibitem{9585309}
X.~Chen, C.~Deng, B.~Zhou, H.~Zhang, G.~Yang, and S.~Ma, ``High-accuracy {CSI}
  feedback with super-resolution network for massive {MIMO} systems,''
  \emph{IEEE Wireless Communications Letters}, vol.~11, no.~1, pp. 141--145,
  2022.

\bibitem{9419066}
Z.~Cao, W.-T. Shih, J.~Guo, C.-K. Wen, and S.~Jin, ``Lightweight convolutional
  neural networks for {CSI} feedback in massive {MIMO},'' \emph{IEEE
  Communications Letters}, vol.~25, no.~8, pp. 2624--2628, 2021.

\bibitem{9497358}
S.~Ji and M.~Li, ``{CLNet}: Complex input lightweight neural network designed
  for massive {MIMO CSI} feedback,'' \emph{IEEE Wireless Communications
  Letters}, vol.~10, no.~10, pp. 2318--2322, 2021.

\bibitem{9373670}
Z.~Lu, J.~Wang, and J.~Song, ``Binary neural network aided {CSI} feedback in
  massive {MIMO} system,'' \emph{IEEE Wireless Communications Letters},
  vol.~10, no.~6, pp. 1305--1308, 2021.

\bibitem{wang2022deep}
X.~Wang, X.~Hou, L.~Chen, Y.~Kishiyama, and T.~Asai, ``Deep learning-based
  massive {MIMO CSI} acquisition for {5G} evolution and {6G},'' \emph{IEICE
  Transactions on Communications}, vol. 105, no.~12, pp. 1559--1568, 2022.

\bibitem{6737048}
S.~V. {Venkatakrishnan}, C.~A. {Bouman}, and B.~{Wohlberg}, ``Plug-and-play
  priors for model based reconstruction,'' in \emph{2013 IEEE Global Conference
  on Signal and Information Processing}, 2013, pp. 945--948.

\bibitem{9413947}
W.~Chen, D.~Wipf, and M.~Rodrigues, ``Deep learning for linear inverse problems
  using the plug-and-play priors framework,'' in \emph{ICASSP 2021 - 2021 IEEE
  International Conference on Acoustics, Speech and Signal Processing
  (ICASSP)}, 2021, pp. 8098--8102.

\bibitem{9454311}
K.~Zhang, Y.~Li, W.~Zuo, L.~Zhang, L.~Van~Gool, and R.~Timofte, ``Plug-and-play
  image restoration with deep denoiser prior,'' \emph{IEEE Transactions on
  Pattern Analysis and Machine Intelligence}, vol.~44, no.~10, pp. 6360--6376,
  2022.

\bibitem{9325040}
M.~Zhao, X.~Wang, J.~Chen, and W.~Chen, ``A plug-and-play priors framework for
  hyperspectral unmixing,'' \emph{IEEE Transactions on Geoscience and Remote
  Sensing}, vol.~60, pp. 1--13, 2022.

\bibitem{7542195}
S.~Sreehari, S.~V. Venkatakrishnan, B.~Wohlberg, G.~T. Buzzard, L.~F. Drummy,
  J.~P. Simmons, and C.~A. Bouman, ``Plug-and-play priors for bright field
  electron tomography and sparse interpolation,'' \emph{IEEE Transactions on
  Computational Imaging}, vol.~2, no.~4, pp. 408--423, 2016.

\bibitem{ryu2019plug}
E.~Ryu, J.~Liu, S.~Wang, X.~Chen, Z.~Wang, and W.~Yin, ``Plug-and-play methods
  provably converge with properly trained denoisers,'' in \emph{International
  Conference on Machine Learning}, 2019, pp. 5546--5557.

\bibitem{hurault2022proximal}
S.~Hurault, A.~Leclaire, and N.~Papadakis, ``Proximal denoiser for convergent
  plug-and-play optimization with nonconvex regularization,'' \emph{arXiv
  preprint arXiv:2201.13256}, 2022.

\bibitem{7839189}
K.~Zhang, W.~Zuo, Y.~Chen, D.~Meng, and L.~Zhang, ``Beyond a gaussian denoiser:
  Residual learning of deep {CNN} for image denoising,'' \emph{IEEE
  Transactions on Image Processing}, vol.~26, no.~7, pp. 3142--3155, 2017.

\bibitem{8099783}
K.~Zhang, W.~Zuo, S.~Gu, and L.~Zhang, ``Learning deep {CNN} denoiser prior for
  image restoration,'' in \emph{2017 IEEE Conference on Computer Vision and
  Pattern Recognition (CVPR)}, 2017, pp. 2808--2817.

\bibitem{8365806}
K.~Zhang, W.~Zuo, and L.~Zhang, ``{FFDNet}: Toward a fast and flexible solution
  for {CNN}-based image denoising,'' \emph{IEEE Transactions on Image
  Processing}, vol.~27, no.~9, pp. 4608--4622, 2018.

\bibitem{AuaDRiGa}
S.~Jaeckel, L.~Raschkowski, K.~Borner, and L.~Thiele, ``{QuaDRiGa}-quasi
  deterministic radio channel generator, user manual and documentation,''
  \emph{Fraunhofer Heinrich Hertz Institute, Tech. Rep. v2.6.1,}, 2021.

\bibitem{3GPP38901}
3GPP, ``{5G} study on channel model for frequencies from 0.5 to 100 {GHz},''
  \emph{3rd Generation Partnership Project (3GPP), Tech. Rep. 38.901 V16.1.0,},
  2020.

\bibitem{7744574}
S.~H. Chan, X.~Wang, and O.~A. Elgendy, ``Plug-and-play {ADMM } for image
  restoration: Fixed-point convergence and applications,'' \emph{IEEE
  Transactions on Computational Imaging}, vol.~3, no.~1, pp. 84--98, 2017.

\bibitem{8237460}
T.~Meinhardt, M.~Moeller, C.~Hazirbas, and D.~Cremers, ``Learning proximal
  operators: Using denoising networks for regularizing inverse imaging
  problems,'' in \emph{2017 IEEE International Conference on Computer Vision
  (ICCV)}, 2017, pp. 1799--1808.

\bibitem{8168194}
A.~M. Teodoro, J.~M. Bioucas-Dias, and M.~A.~T. Figueiredo, ``Scene-adapted
  plug-and-play algorithm with convergence guarantees,'' in \emph{2017 IEEE
  27th International Workshop on Machine Learning for Signal Processing
  (MLSP)}, 2017, pp. 1--6.

\bibitem{li2009user}
C.~Li, W.~Yin, and Y.~Zhang, ``User' s guide for {TVAL3: TV} minimization by
  augmented lagrangian and alternating direction algorithms,''
  \emph{[Online].Available: http://www.caam.rice.edu/~optimization/L1/TVAL3/}.

\bibitem{guo2020convolutional}
J.~Guo, C.-K. Wen, S.~Jin, and G.~Y. Li, ``Convolutional neural network-based
  multiple-rate compressive sensing for massive mimo csi feedback: Design,
  simulation, and analysis,'' \emph{IEEE Transactions on Wireless
  Communications}, vol.~19, no.~4, pp. 2827--2840, 2020.

\end{thebibliography}
\end{document}